\documentclass[12pt,a4paper]{article}
\usepackage{ifpdf}
\ifpdf
\pdfpageattr {/Group << /S /Transparency /I true /CS /DeviceRGB>>}
\fi
\usepackage{jheppub}
\usepackage{amsmath}
\usepackage{caption}
\usepackage{subcaption}

\def\bR{\mathbb{R}}
\def\bC{\mathbb{C}}

\newcommand{\beq}{\begin{equation}}
\newcommand{\eeq}{\end{equation}}
\newcommand{\bea}{\begin{eqnarray}}
\newcommand{\eea}{\end{eqnarray}}

\newcommand{\no}{\nonumber}

\newcommand{\eps}{\epsilon}

\newcommand{\tA}{\tilde{A}}
\newcommand{\gi}{g^{-1}}

\def\to{\rightarrow}

\begin{document}
\title{\textbf{3D Gravity, Chern-Simons and Higher Spins:  A Mini Introduction}}
\author[a]{K. Surya Kiran,} \ 
\author[a]{Chethan Krishnan,} \
\author[a]{Avinash Raju}
\emailAdd{ksuryakn@gmail.com}
\emailAdd{chethan.krishnan@gmail.com}
\emailAdd{avinashraju777@gmail.com}
\affiliation[a]{Center for High Energy Physics, \\
  Indian Institute of Science, \\ 
  Bangalore - 560012, \ \ India}
\date{}

\begin{abstract}{
These are notes of introductory lectures on (a) elements of 2+1 dimensional gravity, (b) some aspects of its relation to Chern-Simons theory, (c) its generalization to couple higher spins, and (d) cosmic singularity resolution as an application in the context of flat space higher spin theory. A knowledge of the Einstein-Hilbert action, classical non-Abelian gauge theory and some (negotiable amount of) maturity are the only pre-requisites. 
}
\end{abstract}

\maketitle
\newpage

\newpage
\section{Introduction}

{\em This introductory section is a bit more succinct and telegraphic than the rest of the lectures, and its purpose is more to orient than to educate. So a beginning student might find it more useful to start with Section 2 and then come back to the introduction at a later stage. The background required for these lectures is a knowledge of general relativity (upto the field equations, Einstein-Hilbert action and basic solutions like Schwarzschild) and with  classical non-Abelian gauge theory at the level of any of the standard Quantum Field Theory textbooks. The material presented here on 2+1 d gravity is standard, a classic reference is \cite{Steven} (see also \cite{Achucarro:1987vz, Brown:1986nw, Deser:1981wh, Deser:1983tn, Deser:1982vy}). Reviews of higher spin gravity are \cite{Gaberdiel:2012uj,Ammon:2012wc,Perez:2014pya}.}

Finding a quantum theory of gravity is an old and difficult problem with both technical and conceptual challenges. An example of the former is the non renormalizability of general relativity as a perturbative quantum field theory. An example of a more conceptual question is that of the problem of gauge invariant observables in a diffeomorphism invariant theory (like general relativity is). The list of puzzles in quantum gravity goes way beyond these two. Quantizing gravity could mean the quantization of space-time itself, and the meaning of that is not very clear from numerous angles. 

When we have a difficult problem, it is natural to look for easier models which have the same important conceptual features, while at the same time, where a few of the difficulties are avoided. Gravity in 2+1 dimensions is one such model. The theory is relatively simple as it has no propagating degrees of freedom and the gauge constraints have a possibility of being explicitly solved. Hence, the original motivation to look at this theory was related to quantization, and as a toy model for full fledged quantum gravity in 3+1 dimensions. The simplification in 2+1 dimensions can be understood by the observation that in 2+1 dimensions, general relativity can be rewritten as a close-to-trivial gauge theory called Chern-Simons theory with (in one case) a gauge group $SL(2)\times SL(2)$. This is one of the things we will review.

If we believe that string theory with its numerous new degrees of freedom is the correct quantum theory of gravity, the stand-alone quantization of 2+1 dimensional Einstein gravity discussed in the previous paragraph might seem like a not-so central problem. However it turns out that even in string theory the $\mathrm{AdS_3/ CFT_2}$ correspondence shows up in various contexts (especially in holography and black holes), which means that quantum gravity in AdS$_3$ is still of great interest. Another reason for interest in 2+1 dimensions is a fact (getting only recently explored) that in the tensionless limit, string theory becomes a theory of massless higher spins coupled to gravity, and that in 2+1 dimensions there exist examples of such theories which can again be expressed as Chern-Simons theories. Some aspects of this will also be discussed in these lectures.

As an application of the formalism developed, we will consider (a 3-dimensional toy model of) one specific physical problem where quantum gravity is expected to shed some light. This is the problem of cosmological singularities, for which the Big Bang in our past is the quintessential example. We will find examples of three dimensional spacetimes that mimic big-bang like features in 2+1 dimensions and show that when embedded into a higher spin theory, these singularities can often be gauged away, thanks to the extra gauge invariances of the higher spin theory. The fact that higher spin symmetries are bigger gauge invariances than diffeomorphisms can be taken as an indication that the worldsheet gauge invariance of the strings are bigger than usual spacetime gauge invariances. Therefore troublesome (but diffeomorphism invariant) features like horizons and singularities can potentially be gauge artefacts in a higher spin theory.

Numerous papers on higher spin theory in various contexts of some relevance to us are collected in \cite{Gutperle:2011kf, Ammon:2011nk, Gaberdiel:2010pz, Gaberdiel:2011zw, Gaberdiel:2010ar, Gary:2014ppa, Grumiller:2014lna, Matulich:2014hea, Fareghbal:2014qga, Ammon:2013hba, David:2012iu, deBoer:2013gz, Henneaux:2013dra, Henneaux:2010xg, Barnich:2011mi, Barnich:2010eb, Barnich:2012aw, Krishnan:2013zya, Campoleoni:1, Campoleoni:2, Bagchi:2014iea, Bagchi:2014ysa, Bagchi:2013qva, Bagchi:2012xr, deBoer:2013vca, deBoer:2014fra, Riegler:2014bia, Krishnan:2014wya}.

\section{2+1 Dimensional Gravity} 
Einstein-Hilbert action for gravity coupled to matter in $2+1$ dimensions is given by \cite{Steven}
\beq
I = \frac{1}{16 \pi G }\int_{M} d^3 x {\sqrt{-g}} (R - 2 \Lambda) + I_{matter}. \label{act}
\eeq 
\subsection{Dimensional Analysis}

We work with natural units, $ \hbar = c = 1$. Because $[dx^\mu]=[ds]=L$ in this choice, from $ ds^{2} = g_{\mu \nu} dx^{\mu} dx^{\nu} $, it is clear that $[g_{\mu \nu}] = 1$ and hence $[g] = 1$. Now, the expression for the Riemann tensor in terms of metric tensor tells us that $[R] = L^{-2}$. For $I$ to be dimensionless we have $[G] = L=M^{-1}$, hence this theory is therefore power-counting non-renormalizable. From $\eqref{act}$, $[\Lambda] = L^{-2}$, giving us a dimensionless quantity in the theory, 
\beq
l \sim \frac{1}{ G |\Lambda |^{\frac{1}{2}}}.
\eeq 

\subsection{Equations of motion}
Equations of motion for the action \eqref{act} are, 
\beq
R_{\mu \nu} + \Lambda g_{\mu \nu} - \frac{1}{2} g_{\mu \nu} R = -8 \pi G T_{\mu \nu} \label{eom}
\eeq 
these are diffeomorphism covariant. The number of independent components of Ricci and Riemann tensors in $d$ dimensions are 
\[ \frac{d(d+1)}{2} \quad \& \quad \frac{d(d-1)}{4} \left( \frac{d(d-1)}{2} +1 \right) \]
\noindent respectively. In $2+1$ dimensions it is interesting to note that both these tensors have six independent components. Hence, Riemann tensor can be written completely in terms of Ricci tensor and vice versa. Using this and the symmetries of Riemann tensor, it is easy to show that 
\beq
R_{\mu \nu \rho \sigma} = g_{\mu \rho} R_{\nu \sigma} + g_{\nu \sigma} R_{\mu \rho} - g_{\mu \sigma} R_{\nu \rho} - g_{\nu \rho} R_{\mu \sigma} - \frac{1}{2} (g_{\mu \rho} g_{\nu \sigma} - g_{\mu \sigma} g_{\nu \rho}) R \label{rmn}.
\eeq 
There is no traceless part, i.e. Weyl curvature tensor is zero. The above equation also means that in vacuum($T_{\mu \nu}=0$), the solutions of Einstein equation are flat for $\Lambda = 0$, and for $\Lambda \neq 0$ they have constant curvature (\textit{i.e.} the Ricci scalar is a constant). That  is, in vacuum when $\Lambda = 0 $, \eqref{eom} becomes
\beq
R_{\mu \nu} = \frac{1}{2} g_{\mu \nu} R 
\eeq 
which upon taking trace implies $R=0=R_{\mu \nu}$ and hence from \eqref{rmn}, Riemann tensor vanishes and solution is locally flat. Similarly, when $\Lambda \neq 0$ vacuum solutions of Einstein equation have constant curvature. This means that $2+1$ dimensional space-time does not have local degrees of freedom. It has curvature only where there is matter, and there are no gravitational waves. 

The fact that there are no local degrees of freedom in this case can also be seen by looking at the number of independent parameters in the phase space of GR. The independent parameters we have here are independent components of spatial metric on a constant time hypersurface, which is $d(d-1)/2$ for GR in $d$ dimensions, and their time derivatives(conjugate momenta) which are again $d(d-1)/2$ in number. Einstein field equations act as $d$ constraints on initial conditions and further, coordinate choice eliminates $d$ degrees of freedom. This leaves us with $d(d-1) - 2d = d(d-3)$ degrees of freedom, which is zero for $d=3$. 

As 2+1 gravity has no propagating degrees of freedom, it has an interesting Newtonian limit. In this limit, we can show that geodesic equation reduces to \cite{Steven}
\beq
\frac{d^2 x_i}{dt^2} + 2 \left(\frac{d-3}{d-2} \right) \partial_{i} \Phi = 0 
\eeq
i.e. in $d=3$ gravity, static masses do not experience Newtonian gravitational force.
\section{First order formalism}
Let us start the discussion with general number of dimensions, $d$. Basic variables in first order formalism are vielbeins $e^{a}_{\mu}$, they are the transformation matrices between tangent space and coordinate frame. If we take the basis vectors of local tangent space to space-time to be orthonormal with Minkowski metric, vielbeins satisfy 
\begin{align}
g^{\mu \nu} e^{a}_{\mu} e^{b}_{\nu} &= \eta^{ab} \no \\
\eta_{ab} e^{a}_{\mu} e^{b}_{\nu} &= g_{\mu \nu}.
\end{align}
Here, Greek indices are spacetime (``world") indices and Latin indices are vielbein (or triad as they are called, in 2+1 dimensions) indices. Collection of all possible vielbeins at every point on $M$ is together called a frame/vielbein bundle. Now, we can work with $V^a = V^\mu e^a_\mu$ instead of $V^\mu$. Covariant derivative of $V^a$ would be, 
\beq
D_\mu V^a = \partial_\mu V^a + \omega^a_{\mu b} V^b \label{cov}
\eeq
where, $\omega^a_{\mu b}$ is a connection in vielbein basis, it is called  the spin connection. The choice of $\omega^a_{\mu b}$ can be fixed by demanding the net parallel transport of $e^a_\mu$ to give a vanishing covariant derivative (see \cite{ortín2007gravity} or section (12.1) of \cite{green1987superstring} for a clear discussion)
\beq
D_{\mu} e_{\nu}^a = \partial_\mu e_\nu^a - \Gamma^\rho_{\mu \nu} e^a_\rho + \eps^{abc} \omega_{\mu b}e_{\nu c} = 0.
\eeq
If the connection $\Gamma^{\rho}_{\mu \nu}$ is torsion free, then
\beq
T^a = D_{\omega} e^{a} = \mathrm{d} e^{a} + \omega^{a}_{b} \wedge e^{b} = 0 \label{c1}
\eeq
where $e^{a} = e^{a}_{\nu} dx^\nu$ is frame 1-form and $\omega^{a}_{b} = \omega^{a}_{\mu b} dx^\mu$ is spin connection 1-form. Equation \eqref{c1} is called Cartan's first structure equation. For torsion free case, expression for $\omega^{a}_{\mu b}$ can be written explicitly in terms of frame 1-forms by inverting them, we will see this for 2+1 dimensional case later. 

The curvature tensor can be defined using the usual expression for gauge field strength, adapted to the present case \cite{ortín2007gravity}
\begin{align}
[D_\mu, D_\nu]V^a = R_{\mu \nu b}^aV^b .
\end{align}
Using \eqref{cov} Riemann tensor then takes the form,
\begin{align}
dx^\mu \wedge dx^\nu R_{\mu \nu b}^a &= (\partial_{[\mu}\omega_{\nu] a}^b - \omega_{[\mu | a}^c \omega_{| \nu ] c}^b) dx^\mu \wedge dx^\nu \no \\
&= d \omega^{b}_{a} + \omega^{b}_{c} \wedge \omega^{c}_{a}
\end{align}
which is analogous to the familiar gauge theory expression
\beq
F = dA + A\wedge A .
\eeq
We can now use this along with metric and spin connection to write Einstein action in first order formalism, 
\beq
I = k \int  \left[  \eps_{a_1 a_2 \dots a_D} R^{a_1 a_2} \wedge e^{a_3} \wedge \dots e^{a_D} + 
\frac{\Lambda}{D ! } \eps_{a_1 a_2 \dots a_D} e^{a_1} \wedge e^{a_2} \dots \wedge e^{a_D} \right] \label{foeh} . 
\eeq
where $R^{a_1 a_2}$ is a curvature two form i.e., $R^{a_1 a_2}\equiv R^{a_1 a_2 a_3 a_4}e_{a_3}\wedge e_{a_4}$.
Varying this action with respect to $\omega$ gives, 
\beq
\mathrm{d}e + \omega \wedge e = 0.
\eeq
This is the torsion free condition. Varying \eqref{foeh} with respect to $e$ gives 
\beq
\mathrm{d} \omega + \omega \wedge \omega = 0 
\eeq
this is a condition that Ricci curvature vanishes(which is Einstein's equation without matter(for $\Lambda = 0$)). \\
\\

\subsection{$2+1$ D} 
In 2+1 dimensions vielbeins and spin connection can be written as one forms, 
\begin{align}
e^a &= e^a_\mu dx^\mu, \quad \omega^a = \frac{1}{2} \eps^{abc} \omega_{\mu bc} dx^\mu
\end{align}
where 3d Levi-Civita tensor $\eps^{abc}$ is an invariant tensor of $SO(3,1)$(we use the convention $\epsilon^{012}= 1$). The existence of this object is the crucial reason why it is natural to write 2+1 gravity as a Chern-Simons theory, as we will see. 

Einstein-Hilbert action in three dimensions can be written as
\beq
I = \frac{1}{8\pi G} \int \left[e^a \wedge (\mathrm{d}\omega_a + \frac{1}{2} \eps_{abc} \omega^b \wedge \omega^c) + 
\frac{\Lambda}{6} \eps_{abc} e^a \wedge e^b \wedge e^c \right] \label{action2} . 
\eeq
One of the equations of motion is obtained by varying $\omega_a$: 
\beq
T_a = \mathrm{d}e_a + \eps_{abc} \omega^b \wedge e^c = 0 \label{eqn1}. 
\eeq
If triads $e^a_\mu$ are invertible, \eqref{eqn1} can be solved to obtain the following expression for spin connection, 
\beq
\omega^a_\mu = \eps^{abc}e^\nu_c(\partial_\mu e_{\nu b}-\partial_\nu e_{\mu b}) -\frac{1}{2} \eps^{bcd}(e^\nu_b e^\rho_c \partial_\rho e_{\nu d}) e_\mu^a \label{eqn2}
\eeq
invertibility of triad is important as the solution \eqref{eqn2} is a second order equation while \eqref{eqn1} is a first order equation. Non-invertible triads can be important in the quantum theory \footnote{The quantum theory can change depending on the allowed fields one integrates over in the path integral.}, but we will only deal with classical theories.

Varying the action with respect to $e^a$ gives, 
\beq
\mathrm{d}\omega_a + \frac{1}{2} \eps_{abc} \omega^b \wedge \omega^c + \frac{\Lambda}{2} \eps_{abc} e^b \wedge e^c = 0
\eeq
i.e., 
\beq
R_a = \mathrm{d}\omega_a + \frac{1}{2} \eps_{abc} \omega^b \wedge \omega^c = - \frac{\Lambda}{2} \eps_{abc} e^b \wedge e^c 
\eeq
\noindent which is the Einstein's equation in vielbein-spin-connection language. 

Up to boundary terms, action \eqref{action2} is invariant under two sets of gauge symmetries, they are (a) Local Lorentz Transformations (LLT),
\begin{align}
\delta_l e^a &= \eps^{abc} e_b \tau_c \no \\
\delta_l \omega^a &= d\tau^a + \eps^{abc} \omega_b \tau_c \label{local}
\end{align}
where $\tau_a$ is a local function, and (b) Local Translations (LT), 
\begin{align}
\delta_t e^a &= d\rho^a + \eps^{abc} \omega_b \rho_c \no \\
\delta_t \omega^a &= -\Lambda \eps^{abc} e_b \rho_c. \label{trans}
\end{align}
(The subscripts $t$ and $l$ above on $\delta$ are only labels.) These are called local Lorentz transformations and local translations because the number of components of $\tau$ and $\rho$ are precisely equal to the number of parameters of Lorentz transformations ($\frac{d(d-1)}{2}$ in $d$ dimensions) and translations ($d$ in $d$ dimensions) respectively \footnote{However the expressions \eqref{local} and \eqref{trans} have been written specifically for 3 dimensions.}. 

The Einstein-Hilbert action in the second order (\emph{i.e.}, in the metric) formulation is invariant under space-time diffeomorphisms. We now argue that this is a consequence of LLT and LT in the first order (\emph{i.e.}, vielbein plus spin-connection) formulation. Using the identity,
\beq
L_\xi \sigma = d(\xi \cdot \sigma) + \xi \cdot d\sigma
\eeq
for Lie derivative $L_\xi$($\xi$ being a world vector) of a one form $\sigma$, we have 
\begin{align}
L_\xi e^a &= d(\xi \cdot e^a) + \eps^{abc} \omega_b (\xi \cdot e_c) + \eps^{abc} e_b (\xi \cdot \omega_c) + \alpha (E.O.M.) \no \\
L_\xi \omega^a &= d(\xi \cdot \omega^a) + \eps^{abc} \omega_b (\xi \cdot \omega_c) - \Lambda \eps^{abc} e_b (\xi \cdot \rho_c) + \beta (E.O.M.) \label{lie}
\end{align}
$\alpha$ and $\beta$ are functions that depend on $\xi$. Using \eqref{local} and \eqref{trans}, \eqref{lie} can be written as
\begin{align}
L_\xi e^a &= \delta_t e^a|_{\rho^a = \xi \cdot e^a} + \delta_l e^a|_{\tau_c = \xi \cdot \omega_c} + \alpha (E.O.M.) \no \\
L_\xi \omega^a &= \delta_l \omega^a|_{\tau^a = \xi \cdot \omega^a} + \delta_t \omega^a|_{\rho_c = \xi \cdot e_c} + \beta (E.O.M.) 
\end{align} 
thus, space-time diffeomorphisms (thought of as generated by the $\xi$) are not an independent  gauge symmetry. It is a combination of local Lorentz transformations and local translations with parameters $\rho^a = \xi \cdot e^a$ and $\tau^a = \xi \cdot \omega^a$ (This is discussed in \cite{Witten:1988hc}). This relationship is valid when triads are invertible, also, it is clear that the above equivalence holds only for diffeomorphisms that can be built from infinitesimal transformations, i.e. small diffeomorphisms which have an interpretation as being generated by vector fields $\xi$. Large diffeomorphisms that can not be built from infinitesimal transformations should be treated separately, they are important for the quantum theory. 

One of the reasons why quantization of $2+1$ gravity is relatively straightforward is, that the complicated diffeomorphism group can be written in terms of much simpler pointwise gauge transformations as above \cite{Steven}. In fact, this is not unique to 2+1 gravity, but it is a property of topological field theories. 

\section{Connection to Chern Simons theory}
Gravity in 2+1 dimensions behaves like a gauge theory in many ways because its first order action \eqref{action2} is that of a gauge theory: the so-called Chern-Simons theory. 

Let us demonstrate this by taking $A = A^a_\mu T_a dx^\mu$ to be a connection one form of group $G$ on a 3-manifold $M$, i.e $A$ is the vector potential of gauge theory whose gauge group is $G$, the generators of whose Lie algebra are $T_a$. Chern-Simons action for $A$ is then 
\beq
I_{CS} [A] = \frac{k}{4\pi} \int_M Tr \left[A \wedge dA + \frac{2}{3} A \wedge A \wedge A \right] \label{action3}
\eeq
Here $k$ is coupling constant and $Tr$ is the non-degenerate invariant bilinear form on Lie algebra of G, which we will define more concretely below for the various cases. Euler-Lagrange equations of motion of \eqref{action3} are
\beq
F[A] = dA + A \wedge A = 0 \label{flat}
\eeq
hence, $A$ is a flat connection(\emph{i.e.} field strength of $A$ vanishes). This does not mean that $A$ is always trivial, as potential with vanishing field strength might give rise to Aharanov-Bohm effect(as we are working with non-abelian theories throughout, it is understood that when we refer to Aharanov-Bohm, we mean a non-abelian Aharanov-Bohm). 

Now let G be the Poincare group $ISO(2,1)$, let $J^a$ denote generator of Lorentz transformation and $P^a$ that of translations. We then have 
\begin{align}
\left[J^a ,J^b \right] &= \eps^{abc}J_c \no \\
\left[ J^a,P^b \right] &= \eps^{abc}P_c \no \\
\left[ P^a,P^b \right] &= 0. 
\end{align}
If we define the group's invariant bilinear form \footnote{There exists a more standard bilinear form \cite{Witten:1988hc} but that is degenerate once we demand that it commutes with $P^a$.} via
\begin{align}
Tr(J^a P^b ) &= \eta^{ab} \\
Tr( J^a J^b ) &= Tr( P^a P^b ) = 0
\end{align}
and write the connection one form as
\beq
A = e^a P_a + \omega^a J_a 
\eeq
then up to possible boundary terms, it can be checked that the Chern-Simons action \eqref{action3} is same as Einstein-Hilbert action \eqref{action2} with $\Lambda = 0$ and 
\beq
k = \frac{1}{4 \mathrm{G}}
\eeq
Also, it can be checked that the infinitesimal local Lorentz transformations and translations are same as the infinitesimal version of ordinary $ISO(2,1)$ gauge transformation of $A$. 
\beq
A = g^{-1} \bar{A} g + g^{-1}dg 
\eeq

A similar construction can be followed when $\Lambda \neq 0$. First let us look at $\Lambda = - 1/l^2 < 0$ , the $\mathrm{AdS_3}$ space. Take
\begin{align}
A &= \left( \omega + \frac{e}{l} \right) \\
\tilde{A} &= \left( \omega - \frac{e}{l} \right)
\end{align}
where
\beq
A = A^a_\mu T_a dx^\mu = \left( \omega^a_\mu + \frac{e^a_\mu}{l} \right) T_a dx^\mu = \left( \omega + \frac{e}{l} \right)
\eeq
similarly for $\tilde{A}$. These together constitute a connection one form of $SL(2,\bR) \times SL(2,\bR)$ if $T_a$ are the generators of $SL(2,\bR)$ with group algebra, 
\beq
\left[T_a,T_b \right] = \eps_{abc}T^{c}. \label{alg1}
\eeq
By defining the invariant bilinear form 
\beq
Tr \left(T_a T_b \right) = \frac{1}{2}\eta_{ab} \label{IB}
\eeq
we can see that the Chern-Simons action, 
\beq
I[A,\tilde{A}] = I_{CS}[A] - I_{CS}[\tilde{A}]
\eeq
is same as first order action \eqref{action2} with $k = l/4G$, upto boundary terms. 

When $\Lambda = - 1/l^2 > 0$, we will be looking at de-sitter gravity. The generators are that of $SL(2,\bC)$, 
\beq
\left[T_a,T_b \right] = \eps_{abc}T^{c} \label{alg2}
\eeq
with invariant bilinear form $$Tr \left(T_a T_b \right) = \frac{1}{2} \eta_{ab}$$ 
and the connection one forms are 
\begin{align}
A &= \left( \omega + \frac{i}{l} e \right) \\
\tilde{A} &= \left( \omega - \frac{i}{l}e \right).
\end{align}
with \eqref{IB} as an invariant bilinear form and $k= -il/4G$, the Chern-Simons action $$I[A,\tilde{A}] = I_{CS}[A] - I_{CS}[\tilde{A}]$$ is same as first order action \eqref{action2}. Note that $\tilde{A}_aT^a = A^\ast_a T^a$(to make action real). 
This condition also tells us that, unlike in previous cases, $A$ and $\tilde{A}$ are not independent of one another, hence, the gauge group of the theory is $SL(2,\bC)$ and not $SL(2,\bC) \times SL(2,\bC)$. As $F[A]$ of \eqref{flat} is the only gauge covariant local object, hence flat connection implies no local observables.

The algebras \eqref{alg1} and \eqref{alg2} are identical: they are the $SL(2)$ algebra. The coefficients which the field takes values in decides weather it is $SL(2,\bR) \times SL(2,\bR)$ or $SL(2,\bC)$.

One comment: one might worry that writing down theories with non-compact gauge groups is a recipe for trouble. In the quantum theory this leads to non-unitarity due to negative norm states propagating locally. The way our theories will bypass this problem is rather trivially: they have no local degrees of freedom at all, so there is nothing to propagate.

\subsection{Parallel transport on a vector bundle}
Let us now take a small detour \cite{Steven} and look at how to define parallel transport on the vector bundle of triads. We will look at this because it gives us an expression for holonomy, which we will use later.

We now consider parallel transporting of vectors in vielbein basis. Let $x^\mu = x^\mu(s)$ be the curve on the total space $M$, along which we are parallel transporting the vector $v^i$ which is at initial point $x^\mu(0)$. If connection here is $A_\mu^a$, then the parallel transport equation is \textit{defined} \footnote{Note that the discussion on parallel transport is essentially identical to the standard discussion on parallel transport in Riemannian differential geometry (for example), but without the restriction that the connections we are using have to be Christoffel symbols. The connection coefficients can have legs on some arbitrary fiber, that is, $A_\mu^a T^i_{a j}$ stands for a connection $\Gamma^i_{\mu j}$. } as \cite{Steven}
\beq
\frac{dv^i}{ds}+\frac{dx^\mu}{ds}A_\mu^a T^i_{a j} v^j = 0
\eeq
Converting to integral form, we get
\beq
v^i(s) = v^i(0) - \int_0^s ds \frac{dx^\mu(s)}{ds}A_\mu^a T^i_{a j} v^j(s).
\eeq
This can be solved by iteration similar to the Dyson series in QFT
\begin{align}
v^i(s) &= v^i(0) + \int_0^{s_1} ds_1 \frac{dx^\mu(s_1)}{ds_1}A_\mu^a T^i_{a j}\left( v^j(0) + \int_0^{s_2} ds_2 \frac{dx^\mu(s_2)}{ds}A_\mu^a T^j_{a k} (v^k(0)+\dots) \right) \no \\
v^i(s) &= U^i_j(s,0) v^j(0),
\end{align}
where
\beq
U^i_j(s_n,s_m) = P \exp \left( - \int_{s_m}^{s_n} ds \frac{dx^\mu(s)}{ds} A_\mu^a T^i_{a j} \right)
\eeq
in the above expression, $P$ implies path ordering. The above expression can be Taylor expanded in following way, 
\begin{align} 
&P \exp \left( - \int_{s_m}^{s_n} ds \frac{dx^\mu(s)}{ds} A_\mu^a T^i_{a j} \right) = \no \\
&\sum_{n=0}^\infty \frac{(-1)^n}{n !} \int_{0}^{s_1} ds_1 \dots \int_{0}^{s_n} ds_n P \left( \frac{dx^{\mu_1}(s_1)}{ds_1} A_{\mu_1}^a T_{a } \dots \frac{dx^\mu(s_n)}{ds_n} A_\mu^b T_{b } \right).
\end{align}
If $ \alpha : [0,1] \to M$ is a closed curve, then parallel transport matrix $U(1,0)$ is called the gauge holonomy matrix of $\alpha$, or holonomy for short. 

By expanding path ordered exponential and using Stoke's theorem, $U$ matrix for an infinitesimal curve can be shown to be 
\beq
U^i_j \approx \delta^i_j - \int_s F^a_{\mu \nu} T^i_{aj} \Sigma^{\mu \nu} + \dots
\eeq
from this, if geometry does not have non trivial cycles and $F^a_{\mu \nu} = 0$, there are no non-trivial observables. But, if the geometry has non trivial cycles, then the boundary doesn't enclose a complete surface, hence we can not use Stoke's theorem like we did in previous equation. If holonomy does not vanish even when $F^a_{\mu \nu}$ vanishes, it is precisely the Aharanov-Bohm phase shift.
\textbf{Therefore holonomies are the observables associated with a flat connection.} In the classical theory, holonomies can be used to distinguish classical solutions and in quantum theory they are gauge invariant non-local observables.

\subsection{Boundary terms and WZW Action}
In this section, let us look at gauge symmetries and boundary terms that come up during gauge transformation of \eqref{action3}. The Chern-Simons action depends directly on the gauge-variant $A$ and not on the gauge-invariant field strength $F$ in a conventional gauge theory. Let us look at its behaviour under gauge transformation of \eqref{action3}, 
\beq
A = \gi dg + g\tA \gi.
\eeq
Substituting this in action, we get 
\begin{align}
I_{CS}[A] &= I_{CS}[\tA] + \frac{k}{4\pi}\int_{M} Tr \bigg[ (\gi dg)\wedge(d\gi)\wedge(dg) + (dg)\wedge (d\gi)\wedge \tA \\ \nonumber &+ (dg\gi)\wedge d\tA  - (\gi dg)\wedge (\gi \tA)\wedge (dg) + (\gi \tA g)\wedge (d\gi)\wedge (dg) \\ \nonumber &+ \tA \wedge (gd\gi)\wedge \tA - (\gi \tA g)\wedge(\gi \tA)\wedge (dg) \\ \nonumber &+ \frac{2}{3}\bigg( (\gi dg)\wedge(\gi dg)\wedge(\gi dg) + (\gi dg)\wedge(\gi dg)\wedge(\gi \tA g) \\ \nonumber &+ (\gi dg)\wedge(\gi \tA g)\wedge(\gi dg) + (\gi dg)\wedge(\gi \tA g)\wedge(\gi \tA g)\\ \nonumber &+(\gi \tA g)\wedge (\gi dg)\wedge(\gi dg) + (\gi \tA g)\wedge(\gi dg)\wedge(\gi \tA g) \\ \nonumber &+(\gi \tA g)\wedge(\gi  \tA g)\wedge(\gi dg) \bigg) \bigg]
\end{align}
now observe that second and third term in the above expression can be grouped together as $-d(dg\gi)\wedge \tA$. Terms linear and quadratic coming from kinetic and interaction parts cancel each other, leaving us with 
\begin{align}
I_{CS}[A] &= I_{CS}[\tA] - \frac{k}{4\pi}\int_{\partial M}Tr \Big[(dg \gi)\wedge \tA \Big] \\ \nonumber &+ \frac{k}{4\pi}\int_{M} Tr \bigg[ (\gi dg)\wedge(d\gi)\wedge(dg) + \frac{2}{3}(\gi dg)\wedge(\gi dg)\wedge(\gi dg) \bigg]  
\end{align}
the last term upon simplification yields,
\begin{equation}
-\frac{k}{12\pi} \int_{M} Tr \Big[  (\gi dg)\wedge(\gi dg)\wedge(\gi dg)  \Big]
\end{equation}
thus under a finite gauge transformation of the Chern-Simons action, we have
\begin{align*}
I_{CS}[A] = I_{CS}[\tA] &- \frac{k}{4\pi}\int_{\partial M}Tr \Big[ (dg \gi)\wedge \tA \Big] \\ \nonumber &- \frac{k}{12\pi} \int_{M} Tr \Big[  (\gi dg)\wedge(\gi dg)\wedge(\gi dg)  \Big]
\end{align*}
boundary term in RHS vanishes if the space is compact. If group $G$ is also compact, the last term is related to the winding number of the gauge transformation \cite{Nakahara}, its value will be $ 2 \pi n$($n$ is integer) for appropriate $k$. Hence $\exp(iI_{CS}) $ which occurs in path integral is gauge invariant. If $M$ is not closed, then we need to add boundary contributions to \eqref{action3} to make the variational principle well defined(as the expression is for a closed manifold). This can be seen explicitly by varying $A$ in Chern-Simons action \eqref{action3}, 
\beq
\delta I_{CS}[A] = - \frac{k}{4\pi} \int_{\partial M} Tr[A \wedge \delta A] \\ + \alpha (E.O.M.)
\eeq 
where $\alpha (E.O.M.)$ means terms proportional to equations of motion, the other term does not vanish if $M$ is not closed and action wont have an extrema. 

We can look at a simple example of scalar field theory to illustrate a solution for situation at hand. Consider a scalar field $\phi$ whose action is written as
\beq
I[\phi] = \frac{1}{2} \int_{M} d^nx \sqrt{-g} \phi \Delta \phi
\eeq 
where $\Delta$ is a Laplacian $\nabla_\mu \nabla^\mu$. If unit normal at boundary $\partial M$ is $n^\mu$ and $h$ is induced metric on the boundary, varying the above action with respect to $\phi$ gives 
\beq
\delta I[\phi] = \int_{M} d^nx \sqrt{-g} \delta \phi \Delta \phi + \frac{1}{2} \int_{\partial M} d^{n-1}x \sqrt{h}(\phi n^\mu \nabla_\mu \delta \phi - \delta \phi n^\mu \nabla_\mu \phi ). \label{add}
\eeq 
This action clearly has extrema only when both $\phi$ and its normal derivatives vanish at the boundary. Correction term to be added depends on boundary condition. If we take $\phi$ to be fixed(i.e. $\delta \phi =0$) at boundary (Dirichlet boundary condition), the term 
\beq
I_{\partial M} [\phi] = - \frac{1}{2} \int_{\Sigma} d^{n-1}x \sqrt{h} \phi n^\mu \nabla_\mu \phi
\eeq 
cancels the boundary term in \eqref{add}($\Sigma$ is a surface such that $\Sigma + \partial M$ is closed). Total action is then 
\beq
I'[\phi] = I[\phi] + I_{\partial M} [\phi] = - \frac{1}{2} \int_{M} d^{n}x \sqrt{-g} \nabla^\mu \phi \nabla_\mu \phi
\eeq 
If we take the normal derivative to be fixed at boundary (Neumann boundary condition) instead of fixing $\phi$ itself, total action will then be $I''[M] = I[M] - I_{\partial M} [\phi]$. Hence, as the correction term for action to have appropriate extrema depends on boundary condition, so does the complete action. This can be expected because boundary contribution depends on the behaviour of the field at the boundary. 

A similar simple exercise can be done with Chern Simons action, if we choose a complex structure and look at AdS theory \cite{Campoleoni:1}, 
\begin{align}
\partial M &= R \times S^1 \no \\
x^\pm &= \frac{t}{l} \pm \phi.
\end{align} 
We then have
\begin{align}
\delta I_{CS}[A] &= - \frac{k}{4\pi} \int_{\partial M} Tr[A \wedge \delta A] \no \\
&= - \frac{k}{4\pi} \int_{R \times S^1} dx^+ dx^-Tr[A_+ \delta A_- - A_- \delta A_+ ] 
\end{align} 
apart from cases like $A_+ = 0$ or$A_- = 0$ at boundary(which can also be worthy of study), to define a general boundary value problem, we can add
\beq
I_{\partial M} [A] = \frac{k}{2\pi} \int_{M} dx^+ dx^- Tr[A_+ A_-].
\eeq 
Depending on weather $A_+$ is held constant or $A_-$, final action will be
\beq
I_{CS}'[A] = I_{CS}[A] \pm I_{\partial M} [A]. 
\eeq 
If we keep $A_+$ fixed, the modified Chern Simons action $I_{CS}'[A] = I_{CS}[A] + I_{\partial M} [A]$ transforms under gauge transformation $A = g^{-1} dg + g^{-1}\tilde{A}g$ as 
\beq
I_{CS}'[A] = I_{CS}'[\tilde{A}] + k I^+_{WZW} [g,\tilde{A}_+]. 
\eeq 
where $I^+_{WZW} [g,\tilde{A}_+]$ is the action of a chiral Wess-Zumino-Witten model on the boundary $\partial M$,
\beq
I^+_{WZW} [g,\tilde{A}_+] = \frac{1}{4\pi} \int_{\partial M}Tr \left[ g^{-1}\partial_+g g^{-1}\partial_-g - 2 g^{-1}A_-g \tilde{A}_+ \right] + \frac{1}{12\pi} \int_M Tr \left[ g^{-1}dg \right]^3 
\eeq 

This implies that number of physical degrees of freedom of Chern-Simons theory, 2+1 gravity in particular, depends on whether space time has a boundary. If there is boundary, gauge invariance is broken on it and these gauge degrees of freedom are dynamical, each broken symmetry adding infinite dimensional space of solutions that are not equivalent. 
\section{Chern-Simons Higher Spin \textbf{$\mathrm{AdS_3}$}}
We have previously seen Chern-Simons $\mathrm{AdS_3}$ theory. Now, let us generalize the theory to include more degrees of freedom. This can be done by increasing the rank of Chern-Simons gauge group: the resulting theory is called a higher spin Chern-Simons theory. We will only discuss $\mathrm{AdS_3}$ case in detail, but similar statements exist for flat space and $\mathrm{dS_3}$ as well. 

As an example, let us promote the gauge group of $\mathrm{AdS_3}$ to $SL(3,\bR) \times SL(3,\bR)$. This can be done by introducing five symmetric and traceless generators $T_{ab}$ to the three generators $T_a$ of $SL(2,\bR)$ to give eight generators of $SL(3,\bR)$. Their algebra is  
\begin{align}
[T_a,T_b] &= \eta_{ab} \no \\
[T_a,T_{bc}] &= \eps^d_{a(b}T_{c)d} \no \\
[T_{ab},T_{cd}] &= -(\eta_{a(c}\eps_{d)be} + \eta_{b(c}\eps_{d)ae} ) T^e
\end{align} 
where $(ab)$ means the symmetric product in indices $a$ and $b$. 

If we generalize the frame fields and connections to generators $T_{ab}$ as $e^{ab}_\mu$ and $\omega^{ab}_\mu$ respectively, we can try defining the $SL(3,\bR)$ connections as
\begin{align}
A &:= \left( \omega^a_\mu + \frac{e^a_\mu}{l} \right) T_a dx^\mu + \left( \omega^{bc}_\mu + \frac{e^{bc}_\mu}{l} \right) T_{bc} dx^\mu \equiv \left( \omega + \frac{e}{l} \right) \no \\
\tilde{A} &:= \left( \omega^a_\mu - \frac{e^a_\mu}{l} \right) T_a dx^\mu + \left( \omega^{bc}_\mu - \frac{e^{bc}_\mu}{l} \right) T_{bc} dx^\mu \equiv \left( \omega - \frac{e}{l} \right).
\end{align} 
By defining the invariant bilinear form to be 
\begin{align}
Tr(T_aT_b) &= 2 \eta_{ab} \no \\
Tr(T_aT_{bc}) &= 0 \no \\
Tr(T_{ab}T_{cd}) &= -\frac{4}{3}\eta_{ab}\eta_{cd} + 2(\eta_{ac}\eta_{bd} + \eta_{ad}\eta_{bc} )
\end{align} 
and taking $k = l/4G$, action $$I[A,\tilde{A}] = I_{CS}[A] - I_{CS}[\tilde{A}]$$ will become 
\beq
I = \frac{1}{8\pi G} \int \left[e \wedge R + \frac{1}{3l^2} e \wedge e \wedge e \right] \label{action4} . 
\eeq
which is same as first order action \eqref{action2}, except that fields and connections also include those generated by $T_{ab}$. It turns out that this will give us 2+1 gravity coupled to spin-3 theory\cite{Campoleoni:1,Campoleoni:2}.
By index counting, we can define metric and spin field as \cite{Campoleoni:1}
\begin{align}
g_{\mu \nu} &= \frac{1}{2} Tr[e_{(\mu} e_{\nu)}] \no \\
\Phi_{\mu \nu \rho} &= \frac{1}{9}Tr[e_{(\mu} e_\nu e_{\rho)}]
\end{align} 
where $e_\mu = e^a_\mu T_a + e^{bc}_\mu T_{bc} $. 

We can also work in a more convenient basis for generator matrices, we label the generators as $L_i$(i= -1,0,1) and $W_m$(m = -2,-1,0,1,2). Their algebra takes the form 
\begin{align}
[L_i,L_j] &= (i-j) W_{i+j} \no \\
[L_i,W_m] &= (2i-m)W_{i+m} \no \\
[W_{m},W_{n}] &= -\frac{1}{3}(m-n)(2m^2 + 2n^2 - mn - 8 ) L_{m+n}
\end{align} 
and the invariant bilinear form becomes
\begin{align} 
Tr(L_0L_0) &= 2, &Tr(L_1L_{-1}) = -4, \no \\
Tr(W_0W_0) &= \frac{8}{3}, & Tr(W_1W_{-1}) = -4, \no \\
Tr(W_2W_{-2}) &= 16. \label{IVB}
\end{align} 
In this new basis, connections can then be written as 
\begin{align} 
A &= A^aL_a + A^mW_m \no \\
\tilde{A} &= \tilde{A}^aL_a + \tilde{A}^mW_m .
\end{align} 
Relation between the two representations and explicit matrices of generators in fundamental representation are given in Appendix. Parallel discussion follows for $\mathrm{dS_3}$ space, it is discussed in \cite{Krishnan:2013cra}. Aspects of flat space theory can be found in \cite{Bagchi, Avinash, Troncoso}

\section{Singularity Resolution using Chern-Simons Theory}
In this section, we will demonstrate that higher Spin Chern-Simons theory can be used to resolve singularities. String theory is expected to resolve various space-time singularities. But it is not straightforward to resolve singularities in cosmological space-time as they are time dependent: string quantization is usually possible only on time independent supersymmetric backgrounds. One way to go about it is to look at cosmological quotients of flat space. As the covering space is flat, we can still use string theory to explore these singularities. Some work in this direction has been done on Milne orbifold(orbifold obtained by quotienting flat space with a boost). In this case, it turns out that some tree level string scattering amplitudes are singular and hence, string theory breaks down \cite{Berkooz:2002je}. 

We will consider the Milne singularity in the context of Chern-Simons higher spin theory in 2+1 dimensional flat space. The tensionless limit of string theory is expected to be captured by higher spin theories. Hence, in this limit, we can think heuristically that the world sheet gauge symmetries of tree level string theory are realized as gauge symmetries of classical higher spin theory. So in the tensionless limit we can ask whether space-time singularities are gauge artefacts, and if they are, can we get rid of them by doing a gauge transformation. In fact, we will now show that we can remove the singularity in the Milne Orbifold by doing a flat space higher spin gauge transformation.  Therefore, instead of calling it a singularity resolution, we can say that we are getting rid of singularity by doing a gauge transformation \cite{Krishnan:2013tza}.

In the previous section, we have discussed Chern-Simons higher spin $AdS_3$ theory. It turns out that \cite{Avinash} this can be translated to a flat space theory by making a substitution 
\beq
\frac{1}{l} \to \eps 
\eeq
where $l$ is AdS radius and $\eps$ is Grassmann parameter defined by $\eps^2 = 0$ \footnote{One way to see this is to write down the Chern-Simons action for the $SL(2, R) \times SL(2, R)$ theory, namely (4.14), and note that it reduces to the first order Einstein action (3.12) without a cosmological constant (in other words, flat space Chern-Simons gravity) when we make the replacement $\frac{1}{l} \rightarrow \epsilon$. An analogous translation can be seen to apply for higher rank Chern-Simons theories as well. We refer the reader to \cite{Avinash} for details, and for further evidence that this connection actually goes far beyond a mere map between the actions. The basic reason why this works is because of the fact that flat space higher spin gauge groups are Inonu-Wigner contractions of AdS higher spin gauge groups. In the pure gravity case, this is the familiar statement that $ISO(2,1)$ is Inonu-Wigner contraction of $SL(2,R) \times SL(2,R)$.}. We take the $SL(3)$ matrices that are defined in Appendix. Milne metric in 2+1 dimensions is \cite{Krishnan:2013tza}
\beq
ds^2 = -dT^2 + r_C^2 dX^2 + \alpha^2 T^2 d\phi^2. \label{met}
\eeq
where the parameters $\alpha$ and $r_C$ in terms of Mass (M) and Spin (J) are 
\begin{align}
\alpha &= \sqrt{M}, & r_C = \sqrt{\frac{J^2}{4M}}.
\end{align}
We are setting $8G = 1$. Space time behaves like a double cone and there is a causal singularity at $T=0$ where $\phi$-circle crunches to a point before expanding in a big-bang. Singularities in 3 dimensions are causal structure singularities, not curvature singularities, because the spacetime has constant curvature at every regular point. In particular, since it is a spacelike circle that is shrinking to zero size at $T=0$, what we have is a cosmological singularity, not a horizon. This is discussed in detail elsewhere \cite{Barnich:2012aw,Cornalba1,Cornalba2,Cornalba3}. From \eqref{met}, the triads (vielbein) and spin connection one forms for the Milne universe are 
\begin{align}
e^T &= dT, \, e^X = r_C dX, \, e^\phi = \alpha T d \phi, \\
\omega^T &= 0 = \omega^\phi, \, \omega^X = \alpha d\phi .
\end{align}
The Chern-Simons connection is then 
\begin{align}
A^{\pm} &= (\omega^a \pm \eps e^a)T_a \\
&= \pm(\eps dT)T_T + (\alpha d\phi \pm r_C dX)T_X \pm (\eps \alpha T d\phi)T_\phi
\end{align}

Let us now look at holonomy, the $\phi$-circle holonomy matrix is $\omega^\pm_\phi = 2 \pi \alpha (T_X \pm \eps T T_\phi)$, it has the eigenvalues $(0, \pm 2 \pi \alpha)$. Similarly, $X$-circle holonomy matrix $\omega^\pm_X = \pm 2 \pi r_C T_X$ has the eigenvalues $(0, \pm 2 \pi r_C \eps)$. 

The characteristic polynomial coefficients of these holonomy matrices are captured by
\begin{align}
\Theta_\phi^0 &\equiv det(\omega_\phi) = 0, &\Theta_X^0 &\equiv det(\omega_X) = 0, \no \\
\Theta_\phi^1 &\equiv Tr(\omega^2_\phi) = 8 \pi^2 \alpha^2, &\Theta_X^1 &\equiv Tr(\omega^2_X) = 0. \label{cpc}
\end{align}
the $\pm$ superscript is dropped as the polynomials are identical for both. The higher spin theory that we consider should also have same characteristic polynomial for it to describe the same gauge configuration. Because, two matrices that have same characteristic polynomial ($Det[A-\lambda I]$) have the same eigenvalues. 

From now on, let us drop the superscripts on the Chern-Simons connection and work with the holomorphic component($A^+$), the anti-holomorphic part can be worked out in the same way. Adding the higher spin components, we get 
\beq
A' = A + \sum_{n=-2}^{n=2} (C^n + \eps D^n) W_n
\eeq
where $C^n$ and $D^n$ are frame fields and connection associated with generators $W_n$. For simplicity, let us assume that they depend only on $T$ and are independent of $\phi$ and $X$.  Note that if we can find {\em some} resolution, we can declare victory. Using definition $g_{\mu\nu} = Tr[e_\mu e_\nu]$ and \eqref{IVB}, the metric now transforms into 
\beq
g_{\mu \nu}' = g_{\mu \nu} + \frac{4}{3} D^0_\mu D^0_\nu - 2 D^{1}_\mu D^{-1}_\nu - 2 D^{-1}_\mu D^{1}_\nu + 8 D^{2}_\mu D^{-2}_\nu + 8 D^{-2}_\mu D^{2}_\nu 
\eeq
and the holonomy matrices 
\begin{align}
\omega_\phi &= 2 \pi (\alpha T_X + \eps \alpha T T_\phi + C^n W_n + \eps D^n W_n), \no \\
\omega_X &= 2 \pi (r_C T_X + C^n W_n + \eps D^n W_n)
\end{align}
should have characteristic polynomial coefficients same as \eqref{cpc}. The resulting relations are many for general $C$s and $D$s. Also the flatness condition for the new connection
\beq
F' = dA' + A' \wedge A'
\eeq
adds further constraints via the equations of motion.

As our goal is to resolve the Milne singularity by hook or crook, instead of solving for a general case, we can try and look for a choice of $C$s and $D$s that satisfy all the necessary conditions. First, let $C^m_\mu =0$, the holonomy constraints then give us
\beq
D^0_\phi = 3 (D^2_\phi + D^{-2}_\phi)
\eeq
in this choice, the remaining constraints from equation of motion will be satisfied if we set $D^0_\phi=3D^2_\phi$ and take all the remaining $D$s to vanish. The resultant change in metric is 
\beq
g_{\phi \phi}' = g_{\phi \phi} + 12 (D^2_\phi)^2,
\eeq
all other components will remain the same. The resultant Ricci scalar 
\beq
R = \frac{12 (D^2_\phi)^2 \alpha^2}{(12 (D^2_\phi)^2 + T^2 \alpha^2)^2}
\eeq
is finite and continuous at $T=0$. Of course, since the metric is regular everywhere, this is expected, but it is nice to check the continuity of the curvature scalar at $T=0$ nonetheless. Hence, the shrinking Milne universe now has a minimum radius at $T=0$ and also all the symmetries of the theory are maintained. The metric now is smooth, instead of crunching to a point followed expanding in a big bang. Hence, the Milne universe is desingularized. The non vanishing components of higher spin field $\Phi_{\mu\nu\rho} = \frac{1}{9}Tr[e_{(\mu}e_\nu e_{\rho)}]$ of resolved Milne orbifold
\begin{align}
\Phi_{\phi\phi\phi} &= - \frac{16}{3} (D^2_\phi)^3 + \frac{4}{3} D^2_\phi T^2 \alpha^2 \no \\
\Phi_{\phi\phi X} &= \frac{8}{9}D^2_\phi r_C T \alpha\no \\
\Phi_{\phi X X} &= \frac{4}{9} r_C^2 D^2_\phi,
\end{align}
are regular everywhere and can be thought as matter fields supporting the resolved geometry. Note also that the $T$ coordinate gives the spacetime a natural global time-ordering, so there is no possibility of closed timelike curves. 

\section{Final Comments}
Gravity is much simpler in 2+1 dimensions, it has no local degrees of freedom, and in that sense it can be solved exactly \cite{Witten:1988hc}. It is relatively easy to quantize it because diffeomorphism group can be written as simple pointwise gauge group. In 2+1 dimensions, we can write spin connection as a one form, so, it is then natural to write gravity as a Chern-Simons theory. Finally, if the space is closed, solution space of Chern-Simons theory is finite dimensional, on the other hand, if there is a boundary, breaking of gauge symmetry makes the solution space infinite dimensional. 

Chern-Simons theory also served as a simple way to couple higher spins to gravity in 2+1 dimensions. This, as we discussed in the introduction, is interesting as a toy model for string theory in the tensionless limit. The gauge invariances of string theory become unbroken in the tensionless limit and we expect that they are related to the higher spin gauge symmetries of the higher spin theory: these enhance the ordinary diffeomorphism invariance of general relativity in spacetime. We used this gauge redundancy to gauge away the Milne singularity as a toy model for resolving the big bang. Progress in embedding our singularity resolution in a stringy context has been made in \cite{Craps:2014wpa, Kiran:2014kca}, where the phenomenon was found to be robust. That there exist consistent boundary conditions for our solutions was elucidated in \cite{Gary:2014ppa}. It will be very interesting to understand the manifestations of such a resolution in a symmetry-broken phase of higher spin theory. This could work as an instructive way in which string theory resolves physically realistic cosmological singularities.

\section*{Acknowledgments}
CK would like to thank Oleg Evnin, Gabriele Ferretti and Amir Esmaeil Mosaffa for warm hospitality at their respective institutions and the participants and organizers of the IPM School on AdS/CFT (2014) for questions and discussions that improved the manuscript.

\appendix
\section{Fundamental Matrix Representation}

Generators $T_a$ and $T_{bc}$ can be written in terms of generators in fundamental matrix representation as follows, 
\begin{align} 
T_0 &= \frac{1}{2} (L_1 + L_{-1}), &T_1 = \frac{1}{2} (L_1 - L_{-1}), \no \\
T_2 &= L_0, \no \\ 
T_{00} &= \frac{1}{4} (W_2 + W_{-2} + 2W_0), &T_{01} = \frac{1}{4} (W_2 - W_{-2}), \no \\
T_{11} &= \frac{1}{4} (W_2 + W_{-2} - 2W_0), &T_{02} = \frac{1}{2} (W_1 - W_{-1}), \no \\ 
T_{22} &= W_0, &T_{12} = \frac{1}{2} (W_1 + W_{-1}). 
\end{align} 

Matrices of generators in the fundamental representation are \cite{Castro:2011fm}
\begin{align}
L_1 &=
\begin{bmatrix}
0 & 0 & 0 \\
1 & 0 & 0 \\
0 & 1 & 0
\end{bmatrix}, \, \, \, \, \, \, 
L_0 = 
\begin{bmatrix}
1 & 0 & 0 \\
0 & 0 & 0 \\
0 & 0 & -1
\end{bmatrix}, \, \, \, \, \, \, 
L_{-1} = 
\begin{bmatrix}
0 & -2 & 0 \\
0 & 0 & -2 \\
0 & 0 & 0
\end{bmatrix}, \no \\
W_0 &= \frac{2}{3}
\begin{bmatrix}
1 & 0 & 0 \\
0 & -2 & 0 \\
0 & 0 & 1
\end{bmatrix}, \, \, \, \, \, \, 
W_1=
\begin{bmatrix}
0 & 0 & 0 \\
1 & 0 & 0 \\
0 & -1 & 0
\end{bmatrix}, \, \, \, \, \, \, 
W_2 = 2
\begin{bmatrix}
0 & 0 & 0 \\
0 & 0 & 0 \\
1 & 0 & 0
\end{bmatrix}, \no \\
W_{-2} &= 2
\begin{bmatrix}
0 & 0 & 4 \\
0 & 0 & 0 \\
0 & 0 & 0
\end{bmatrix}, \, \, \, \, \, \, 
W_{-1}=
\begin{bmatrix}
0 & -2 & 0 \\
0 & 0 & 2 \\
0 & 0 & 0
\end{bmatrix}. 
\end{align}

\newpage

\bibliographystyle{JHEP}
\bibliography{TCSTD2}

\end{document}